# Principles for Inconsistency

Shel Finkelstein
SAP Research
3410 Hillview Avenue
Palo Alto, CA 94304
+1 650 461-1741

shel.finkelstein@sap.com

Rainer Brendle
SAP Research
3410 Hillview Avenue
Palo Alto, CA 94304
+1 650 687-4721

rainer.brendle@sap.com

Dean Jacobs
SAP Research
Dietmar-Hopp-Allee 16
69190 Walldorf, Germany
+49 171-3363250

dean.jacobs@sap.com

## ABSTRACT
Data consistency is very desirable because strong semantic properties make it easier to write correct programs that perform as users expect. However, there are good reasons why consistency may have to be weakened to achieve other business goals. In this CIDR 2009 Perspectives paper, we present real-world reasons inconsistency may be necessary, offer principles for managing inconsistency coherently, and describe implementation approaches we are investigating for sustainably scalable systems that offer comprehensible user experiences despite inconsistency.

## Categories and Subject Descriptors
H.2.4 [**Systems**]: Concurrency, Distributed databases, Object-oriented databases, Parallel databases, Query processing, Rule-based databases, Transaction processing. H.2.5 [**Database Administration**]: Logging and recovery. H.2.8 [**Database applications**]. J.1 [**Administrative data processing**]: Business, Financial, Manufacturing. D.2.11 [**Software Architectures**]: Data abstraction, Patterns. D.1.3 [**Concurrent Programming**]: Distributed programming, Parallel programming.

## General Terms
Design, Management, Performance, Reliability.

## Keywords
Consistency, data management, database, transaction processing, availability, manageability, scalability, business applications, business objects, distributed systems, parallelism, concurrency.

## 1. INTRODUCTION
*"A foolish consistency is the hobgoblin of little minds." Ralph Waldo Emerson, "Self-Reliance", 1841 [4]*

A database, or more generally a data management system (DMS), is Consistent if its state satisfies given integrity constraints. A transaction is Consistent if its actions on consistent DMS states always result in consistent states. Atomicity (all-or-nothing), Isolation (transactions behave as if executed in serial order) and Durability (committed changes are never lost), ensure that execution of consistent transactions preserves DMS consistency [7]. Real-world interactions associated with transactions are outside of the DMS, but consistent transactions "normally" behave correctly with respect to these interactions as well, e.g., cash is dispensed only after the cash withdrawal has been posted to your bank account.

For distributed systems with replication, one could define consistency very loosely, with weak integrity constraints, or very tightly, based on single-copy serializability. But weak constraints lose data semantics, while tight constraints eliminate some of the advantages of replication (latency, as well as availability and partition-tolerance, per CAP Principle/Theorem tradeoffs) [1, 6]. Except when there are uncompensatable and unapologizable real-time consequences, it is more useful to speak of Subjective Consistency (performing correct transactions based on the local data state) and Eventual Consistency (convergence to equivalent states at all replicas if there were no further transactions) [3, 9, 10].

Internet-scale data management systems [2,3] must achieve very high levels of scalability and availability, and manage inconsistent replicas in principled ways. There are principles for managing inconsistency of partitioned/replicated data that are application-dependent, and are followed in different ways [2, 3] in internet-scale data management systems. However, principles for managing inconsistency are also relevant for more traditional enterprise applications and databases. This paper describes some of those principles, with particular reference to experiences at SAP, and describes some implementation approaches we are investigating for sustainably scalable systems that offer comprehensible user experiences despite inconsistency.

## 2. PRINCIPLES
This section includes 11 principles for managing DMS inconsistency. Some of these principles are not directly about consistency, but they have consequences for consistency due to the CAP Principle. We hope to generate a vigorous discussion about associated tradeoffs across different schemes for managing replication for availability, such as active systems with <u>asynchronous</u> commits to backups, active systems with <u>synchronous</u> commits to backups, active/active replication with subjective/eventual consistency, and replication with strong consistency.

We do not believe that inconsistency is appropriate for all applications and all data, and we discuss user experience and implementation issues for mixed systems (with different consistency levels) in section 3.

## 2.1 Reality is real
*Business data may not always correctly reflect the state of the world or the business.*

In organizations, people don't have perfect information; the same is true for systems. There is always some delay before entry of information (or before processing of state-changing requests) from devices, humans, partners or other parts of the company. Even without distributed replication, systems must embrace delay as a source of Subjective Consistency [9] that describes system knowledge of reality, rather than reality itself. A DMS should manage latency, incompleteness and even certain kinds of unsoundness, rather than acting as if they don't exist. For example, a business may permit inventory levels to go negative if a packager knows more about current inventory than the system does. Of course uncontrolled integrity constraint violations are problematic; the challenge is to have mechanisms that handle the consequences of inconsistency. For negative inventories, the system should track the history that resulted in negative inventory levels, and eventually account for the discrepancy. See principles 2.2 and 2.9 for related discussions.

## 2.2 Out-of-order works
*Transactions and events sometimes happen in unexpected sequences, temporarily violating integrity constraints.*

One of the simplest constraints is referential integrity, which requires that child entities have valid parent entities. Referential integrity may be handled via a foreign key (reference to a business partner requires existence of the partner) or via a hierarchical data structure (line items must appear in a purchase order). In practice, data might not be received (or even determined) before data that references it. Data generally enters the enterprise IT infrastructure through front-end applications such as Customer Relationship Management (CRM). When entered, it is often incomplete, and since there is little coordination between the users who enter data, often inconsistent. Over time, as users collaborate and business processes are carried out, the data moves towards backend applications and references become more complete. Leads become qualified and turn into Opportunities, which are won and become Orders, which are processed by demand planning, which causes production scheduling, which results in logistics.

Especially in the early stages of the data lifecycle, the DMS should not bureaucratically prevent data entry. Instead, a transaction should be able to enter what's known "now". For example, Opportunities may refer to customers not yet entered. If integrity constraints are violated, warnings can be given, alerts/events generated, and new process steps scheduled. The constraint still exists, but its violations are handled, rather than prevented, so an "inconsistent" business state that would have been regarded as unsound has been transformed into a system-managed exception.

## 2.3 I'll do it eventually
*Secondary data need not be updated with primary data.*

Primary data gets inserted (or perhaps updated) during transactions. Other actions must reliably occur eventually, but may be deferred, not occurring during the original transaction. Examples of such secondary data are aggregates (such as invoice total or quantity on hand) which might be bottlenecks if updated within a transaction, materialized views and OLAP/data warehouses. Eventual update implies that such secondary data will not always be consistent with the primary data. Helland explains why inconsistency of secondary indexes is necessary for highly scalable systems [8], but as he emphasizes (and we'll see in principle 2.5), his point applies beyond secondary indexes.

SAP has utilized similar techniques for many years to avoid database bottlenecks [5]. To reduce user wait times, the SAP transaction model allows a transaction to complete when a descriptor listing pending actions has been committed to the database; the actions themselves are performed after control has returned to the user. Logical locks are held until the actions have completed, but these prevent access by other users, not the user who performed the transaction. If that user immediately queries the state of the system, the result of the transaction may not yet be visible. Applications using asynchronous updates must be designed to tolerate such inconsistency. (The SAP model also supports synchronous updates at commit; that increases response time but avoids this inconsistency.)

Additional sources of inconsistencies due to deferred updates (which we think of as principled procrastination) appear in subsequent principles.

## 2.4 Process steps should focus: At most one transaction per process step
*Processes should be made up of process steps, connected by events. A process step should contain (at most) one transaction, which commits (or at least, attempts to commit) at the end of the step.*

A process, such as transferring an employee from one department to another, should be broken down into a series of steps, such as reassigning the employee's business responsibilities to other employees, that are connected by events. Detailed considerations of the nature of such events, which can be reliable, transactional, or ordered, are beyond the scope of this paper, but aspects are addressed by reliable message queue specifications and products, such as the Java Message Service. For unreliable messaging, at-least-once delivery can be used with idempotence [8].

Identifying process steps with transactions simplifies both the programming model and system management. Including at most one transaction in a process step ensures that the process step can never partially complete its work; it either completes its transaction or it doesn't. And since the transaction boundary is the end of the process step, there is no application-specified work after transaction commit in a process step, so the transaction either committed or it didn't. (System infrastructure will have to determine which if there's a failure.) A committed transaction may enqueue events that result in additional process steps, perhaps specified by application code in the transaction or by system infrastructure. Transaction failures may enqueue post-rollback actions but they must be non-transactional and infrastructure-generated, since the transaction failed and the process step ends at the transaction boundary. Process and processor failures may mean that such post-rollback actions cannot be generated locally, so retry, idempotence and other approaches are necessary.

Not all actions in a process step can (or should) be rolled back. There may be non-transactional writes, e.g., for auditing purposes, which should not be rolled back even if a transaction fails. There may be indirect effects, such as starting or stopping a resource, that must be independently reversed by the system at some appropriate time. Finally, there may be real-world actions taken by a process step, typically post-commit because they won't be rolled back if transaction fails. (Compensation for real-world actions is discussed in principle 2.9.)

## 2.5 Transactions should focus: Only one entity updated per transaction

*Whenever possible, update only a single (frequently hierarchical) entity within a transaction. If updates must involve (or be propagated to) other entities, do so using reliable/transactional queues and process management/eventing techniques.*

An entity is a business object, frequently hierarchical, such as an order and its lineitems. The value of single entity transactions is clearest when they involve data associated with multiple business partners, or even multiple organizational systems running separate databases within the same company. When entities from two different organizational units are accessed in the same transaction, a distributed (two-phase commit) transaction is required, which impacts performance and availability. In the employee transfer example, there is no need to reassign responsibilities across the two departments in the same transaction.

A single organization may partition data by entity type and key, where partitions are managed as separate "serialization units" with separate logs. Entity location is determined dynamically, e.g., by key range partitioning or with a dynamic hash table. Following the focused transaction principle avoids commits across multiple units, which might be distributed commits. Helland proposes that highly scalable systems use single entity transactions for this reason [8].

Even in single database systems, the focused transaction principle may be valuable because it shortens and simplifies transactions, which avoids conflicts, simplifies compensation (principle 2.9), and promotes parallelism and locality. Designing an application from the beginning so that its parts are loosely coupled supports flexible distributed deployment. Ideally, a developer should rely on tight coupling and sequential execution only when it is required for correctness. Of course, good implementations should transparently manage clever performance optimizations as discussed in section 3..

## 2.6 Single Object Update per Process Step: SOUPS on

*Each Process Step consists of (at most) one transaction, updating exactly one data object, possibly also generating reliable and/or transactional events*

This principle is a combination of principles 2.4 and 2.5, Focused Process Steps and Focused Transactions. It's called out as a separate principle because the combined focus dramatically simplifies the programming model and makes local, highly parallel implementation feasible. (Note that enqueue and dequeue operations on event queues, including transactional queues, are always local operations, never distributed transactions, even when delivery is to a remote system.)

Of course, the cost of SOUPS is that applications must tolerate inconsistencies, and the DMS infrastructure must handle inconsistencies. But since we must accept subjective and eventual consistency due to CAP, and there are other sources of inconsistency, an intelligent system design choice is to embrace inconsistency management in a principled way whenever possible.

## 2.7 I remember it well

*Handle (almost all) updates as inserts of new data, and handle deletes by marking data as deleted, rather than actually deleting.*

SAP handles many updates in this "insert-only" way due to regulatory requirements. But this principle can also help with versioning, consistency within a transaction, concurrency control, and, because past descriptions are available, eventual consistency. In the inventory example of principle 2.1, the historical trace might be used to identify a packer as the source of the inconsistency. SAP uses a commutative update strategy ("deltas") and a merge update capability that relate to eventual consistency within a single database system. OLTP transactions should not refer to versions of data explicitly; instead, the DMS should automatically use the appropriate data version.

In practice, unlimited data growth may be an issue, so the DMS should provide data summarization and archival functionality, while still addressing regulatory requirements and eventual consistency.

## 2.8 Beware the consequences

*Data written in transactions should describe what the transactions do, not just transaction consequences. Business processes should handle conflicts or anomalies either immediately (conflict-resolution) or in a deferred mode (data cleansing).*

Describing operations, rather than (just) their consequences, helps support eventual consistency assuming that data is inserted rather than updated (per the previous principle). For example, entering a banking withdrawal means entering the withdrawal, not just the remaining balance. The balance may be updated in the same transaction (if all account transactions are part of an account entity that also includes balance), or perhaps in a deferred transaction (if balance is handled as an aggregate). Note that how transactions read account balance may be affected by this choice. Operations may be described at a high logical level, with consequences happening later; SAP follows this approach when latency is an issue. This principle has a major impact on schema and (combined with other principles, such as non-determinism and deferred update) process design, allowing conflicts to be handled either immediately, via event-handling and process management, or by a deferred cleansing process (which may itself generate events reflecting its actions). It is particularly important for operations on occasionally disconnected devices such as mobile phones.

## 2.9 I think I can
*Process steps and user experience should be designed to support tentative operations and (what Pat Helland [9] aptly calls) apology-oriented computing, where compensation is handled using tentative operations and apologies.*

In SAP's Supply Chain Management (SCM), when one business informs another than a given quantity of an item is Available-To-Purchase at a quoted price by a deadline date/time, that is a business process. The Supplier enters a description of the offer inside its DMS, handling the given quantity as a tentative update of quantity, subject to business rules. A purchase request received by the deadline date will normally be honored, but there may be business reasons (e.g., a disaster at a warehouse) why that can't occur. In either case, the Purchaser will be notified, and appropriate business actions will be taken by both Supplier and Purchaser. The intricate choreography of SCM is designed to handle supply processing, where there are multiple process steps and exception event handling. Such tentative approaches are also suggested by Helland [8].

In Data Management Systems using replication, decisions may be made using subjective consistency (looking at your replica, not across all replicas). Apologies may be required after replicas share information. Users may have been given incorrect information in a consumer transaction ("Your order for the book has been accepted and will be processed") because there were only 5 copies of the book available, and more than 5 were sold. Pat Helland also points out that a warehouse fire could also delay or prevent delivery; as we mentioned in principle 2.1, reality is realer than information systems. In either case, a user may receive an apology indicating that the book will not be delivered. But note the tentativity choreography in book processing introduced by separating Order Entry from Fulfillment; the user has been told that the book order has been received, but not that it will be fulfilled. Overbooking book orders still requires an apology, but the clear separation between Order Entry and Fulfillment makes the user experience more intelligible. Of course, this doesn't help in rare cases such as a warehouse fire or other disaster where a fulfillment promise may have to be abrogated.

There are situations where compensation and apologies may not be possible because of irreversible actions (firing missiles) or real-time constraints (air traffic systems). Inconsistency may not be tolerable in such situations. An intriguing question is whether a single infrastructure can deliver different levels of consistency for different data and different applications, and what data and application management techniques are needed to deliver such capabilities. We discuss this further in section 3.1.

## 2.10 Solipsists get things done quickly
*Each transaction acts based on its local view of the data, without considering other local transactions*

Subjective transactions look at local data without considering operations at other replicas; we propose *solipsistic transactions,* which don't even consider updates occurring locally. Solipsists aren't inconvenienced by pessimistic concurrency control (which can cause waits, timeouts, deadlocks), nor by optimistic concurrency control (which can cause rollback if has data changed since it was read). Instead, solipsistic transactions commit and expect system infrastructure to handle conflicts within a replica, just as that infrastructure handles conflicts across replicas.

Local conflict-resolution may involve composing changes (e.g., using commutative operations, design change integration, or last-update wins), and taking the same compensatory actions (which might include apologies) that would occur if conflicting transactions had addressed different replicas. The crux of this principle is to have a single "end-to-end" conflict-handling mechanism that deals with single and multiple replicas, rather than having different mechanisms for each case.

For focused transactions (principle 2.5) that only update one entity, conflict-resolution may be easier than with a less focused transaction that updates multiple entities. For transactions that generated events, conflict-resolution compensation mechanisms may need to generate other events to compensate for them based on transactional data, metadata and introspection. Transactions that insert only (principle 2.7) and record what they are doing, not just consequences (principle 2.8) are particularly amenable to automated conflict-resolution infrastructure.

## 2.11 The show must go on
*Business services should <u>always</u> be available.*

This is the most important principle; the other principles introduced in this paper help enable it. Processes, processors, disks and network connections may fail; we know that they <u>will</u> sometimes fail. But business transactions and processes should <u>always</u> work, even if/when data is not fully "consistent"—not up-to-date, not completely self-consistent, not reflected in secondary data, and even (sometimes) not what was previously promised.

# 3. IMPLEMENTATION AND USER EXPERIENCE
The previous section listed a series of principles, but made periodic reference to aspects of implementation and user experience. In this section, we summarize those aspects briefly, and talk about additional considerations in both areas.

## 3.1 Implementation considerations
Process steps are connected via events/messages, which may be reliable, or transactional; at-least-once delivery and idempotence can be used with unreliable messaging. Each process step contains at most one transaction, which updates only one entity and generates events. Scheduling for process steps (which may be based on a series of events, not just a single event) is handled by system infrastructure. System infrastructure should be scale-aware code [8], independent of specific applications, but how it behaves for a particular application certainly can depend on the characteristics of the application, its transaction and its data.

This approach supports scalability, parallelism (locally and across partitions and replicas) and fast response times for users, but requires a programming model suitable for infrastructure-based conflict-resolution across multiple applications that use the same entities. Moreover, new applications get added and applications get changed and extended, so a timelessly sustainable application environment must provide both dynamic schema migration and dynamic application migration capabilities, with continuous availability. The infrastructure environment must proscribe admissible changes to schemas and applications; not all changes will be supportable, and only supportable changes can be permitted.

Specifications don't preclude performance optimization by deployment and runtime infrastructure. Infrastructure could collapse steps vertically, turning multiple process steps in the same process into a single sequential process step, and perhaps multiple transactions into a single transaction. Infrastructure could also collapse process steps horizontally, turning multiple transactions for different processes into a single transaction. In either case, that single transaction would have to address local data only. Having small transaction granularity in the programming model allows smart implementations to "right-size" execution to optimize throughput, or trade off throughput for response time.

Because data is (mostly) insert-only and operations (not just consequences) are stored, there can be enough information to handle conflict-resolution, both for solipsistic local transactions and across subjective replicas. One approach we are considering [Hasso Plattner, private communication] involves storing events when they arrive, with inserts treated as events, in a log-structured database (LSDB). What applications view as the current state of the database would be a rollup aggregation of the contents of the LSDB, in the same way that rollforward using a log is an aggregation function. This can be implemented efficiently using main memory database techniques.

Since strong consistency is sometimes necessary for certain data and applications, in section 2.9 we asked whether a "single infrastructure"—an ambiguous term--can deliver different levels of inconsistency for different applications, and what data and application management techniques are needed to deliver such capabilities.

For applications that address the data with different consistency requirements, multiple replicas are required; different replicas may provide different consistencies. For example, a master-slave approach where the master copy handles all updates unapologetically but slaves may have to apologize and compensate might address needs for variegated consistency requirements. We've already seen that subjective Order Entry operations might allow inconsistencies, while Fulfillment might be handled by a master database that provides stronger consistency guarantees. Strong consistency can also be provided using logical locks with coarse granularity, a technique SAP systems use to avoid database bottlenecks [5]. For read-only warehousing requirements, periodic extract from an OLTP system may suffice. But at some level, these approaches integrate separate systems, and at the systems level do not constitute a unified single infrastructure.

Reference data (such as metadata) that seldom changes or is versioned may be managed differently than frequently updated data, and many systems understand that these types of data are different and address them individually. But once again we have two separate technologies that have been integrated, rather than a single infrastructure addressing disparate requirements.

## 3.2 User experience considerations

End-user requirements include ease-of-use (not addressed here), fast response times and no surprises. When users perform business operations, they expect that the effects of those operations are durable and visible. There may be subsequent applications performed by other users that affect the same (or related) data, but there are expected semantics for the composition of those subsequent effects with what the users sees as "my" effects. But the nature of that composition is tricky to capture for updatable data; if I've changed inventory to 15, I recognize that inventory may be 8 or 30 the next time that I examine it. On the other hand, if I'm looking at operations on a bank account, my balance may change, but individual deposits and withdrawals are visible and durable.

So insert-only data (in which balance is an aggregate of deposits and withdrawals) helps deliver a coherent user experience. Even for tentative changes, which might not become permanent (e.g., because a purchaser does not accept an offer from a supplier), the tentative change is visible and durable, but might be marked as obsolete.

Apologies can be difficult for users to address, and some apologies will upset customers, which is not good business, either for a company or for the business application provider. (Sorry that I lost your billion dollar transaction. Sorry that I didn't record your legally necessary business action. Sorry that you've arrived for your vacation, but your hotel reservation was lost because we gave it to someone else.) Building scalable systems that provide good user experience requires defining not only data schemas (with history), application methodologies and systems infrastructure; it also requires defining User Experience so that apologies are comprehensible (and preferably rare). One common approach is to decompose processes into multiple steps, as with the separation of Order Entry, which is visible and durable, from Fulfillment, which may involve races with Fulfillment for other users, and may be impacted by real-world phenomena such as warehouse fires.

Apologies can also be avoided by providing stronger consistency guarantees (trading off other aspects of CAP [1]). Note also that response time for users may degrade due to strong consistency, e.g., when a backup system must receive transaction records before a transaction commits, or when a replica quorum must acknowledge receiving writes before they complete. Hence avoiding apologies can impair a different aspect of user experience, response time.

In section 3.1, we discussed the possibility of having "single infrastructures" than can deliver different levels of consistency and inconsistency. That's a user experience as well as an infrastructure issue; a system that takes business application requirements and automatically delivers appropriate consistency levels based on metadata (describing data, applications, customer expectations, etc.) would be a significant technical and business achievement.

Finally, we believe that knowing your data, your applications and your users allows significant specializations in the design of system infrastructures providing appropriate user experience, including throughput and response time [5].

## 4. CONCLUSION

Some of the inconsistency principles we presented may be controversial; we don't claim that they are universally applicable--consistency is a critical consideration for certain business applications. In this paper, we follow the prescient Emerson by arguing against a foolish (that is, an inappropriate, unnecessary, overly expensive or practically unattainable) consistency.

The papers we reference on internet-scale distributed systems have strongly influenced our views of requirements for high performance and highly parallel scalable data management systems. But experience with SAP applications has also strongly influenced us, and we believe that these inconsistency principles are often sound data management principles for business applications and processes using traditional databases. We welcome feedback and discussion with other researchers exploring related ideas.

*This document contains research concepts from SAP®, and is not intended to be binding upon SAP for any particular course of business, product strategy, and/or development. SAP assumes no responsibility for errors or omissions in this document. SAP does not warrant the accuracy or completeness of the information, text, graphics, links, or other items contained within this material.*

## 5. ACKNOWLEDGMENTS

We'd like to thank many of our colleagues at SAP for insightful discussions relating to topics in this paper, particularly Thomas Heinzel, Hasso Plattner and Heinz Roggenkemper. Discussions by one of the authors with Pat Helland when he was completing reference [9] about object ACID (Associative, Commutative, Idempotent and Distributed) properties also helped motivate writing about data management properties of SAP applications.